\documentclass[preprint]{aastex}
\usepackage{emulateapj5}

\slugcomment{}

\shorttitle{Southern Phase Calibrators}
\shortauthors{Winn, Patnaik, \& Wrobel}

\begin{document}

\title{ Interferometric Phase Calibration Sources\\
in the Declination Range $0\arcdeg$ to $-30\arcdeg$ }

\author{Joshua N.\ Winn\altaffilmark{1,2},
Alok R.\ Patnaik\altaffilmark{3,4},
J.M.\ Wrobel\altaffilmark{5}
}

\altaffiltext{1}{Harvard-Smithsonian Center for Astrophysics, 60
Garden St., Cambridge, MA 02138}

\altaffiltext{2}{NSF Astronomy \& Astrophysics Postdoctoral Fellow}

\altaffiltext{3}{Max-Planck-Institut f\"{u}r Radioastronomie, Auf dem
H\"{u}gel 69, 53121 Bonn, Germany}

\altaffiltext{4}{present address: Apt.\ 1703, 1435 Prince of Wales
Drive, Ottawa K2C 1N5, Canada}

\altaffiltext{5}{National Radio Astronomy Observatory, P.O.\ Box 0,
Socorro, NM 87801}

\begin{abstract}
We present a catalog of 321~compact radio sources in the
declination range $0\arcdeg > \delta > -30\arcdeg$.  The positions of
these sources have been measured with a two--dimensional rms accuracy
of 35~milliarcseconds using the NRAO\footnote{The National Radio
Astronomy Observatory is a facility of the National Science Foundation
operated under cooperative agreement by Associated Universities, Inc.}
Very Large Array.  Each source has a peak flux density $>50$~mJy at
8.4~GHz.  We intend for this catalog to be used mainly for selection
of phase calibration sources for radio interferometers, although
compact radio sources have other scientific uses.
\end{abstract}


\section{INTRODUCTION}
\label{sec:intro}

Radio interferometers measure the amplitudes and phases of Fourier
components of the sky brightness distribution.  The phases are
generally more susceptible than the amplitudes to corruption due to
the electronics and the atmosphere.  Yet the phases encode important
information about the position and structure of the radio source.  For
bright sources, the phases can be corrected by means of
self--calibration \citep{pea84}.  When the source is too faint for
self--calibration, or when the absolute position of the source is
desired, it is essential to correct the phase errors with frequent
observations of a reference source, or ``phase calibrator.''  To be
useful, a phase calibrator should be bright, have an accurately
determined position, and exhibit a simple (preferably compact) radio
structure.  A phase calibrator should also be separated from the
target source by as small a switching angle as possible, to minimize
differential atmospheric phase errors.  This makes it valuable to
identify a dense grid of phase calibrators across the sky.

For historical reasons, the southern sky is not as well--explored for
phase calibrators as the northern sky.  In this paper we help redress
this imbalance by presenting a catalog of 321~compact radio sources
in the declination range $0\arcdeg > \delta > -30\arcdeg$.  The
positions in the catalog have a two-dimensional root-mean-squared
(rms) accuracy of 35~milliarcseconds (mas).  Our catalog is derived
from observations with the NRAO Very Large Array (VLA) that were very
similar in technique to those employed by the Jodrell Bank--VLA
Astrometric Survey \citep[JVAS:][]{jvas1,jvas2,jvas3}.  Our new
observations are described in \S~\ref{sec:observations}.  The details
of data reduction, including the determination of the accuracy of the
derived positions, are given in \S~\ref{sec:data-reduction}.  The
catalog is presented in \S~\ref{sec:catalog}.  In the paper version of
this article, only the first page of the catalog is reproduced; the
entire catalog can be found in the electronic version of this article.
In \S~\ref{sec:summary} we summarize and highlight other possible uses
of the catalog.

\section{OBSERVATIONS}
\label{sec:observations}

In general, the structure of a radio source with a flat spectral index
($\alpha \geq -0.5$, where $S_\nu \propto \nu^{\alpha}$) is dominated
by a core that is compact on milliarcsecond scales.  This was the
underlying assumption of our strategy for selecting potential phase
calibrators, as it was with JVAS.

We selected flat-spectrum sources in the region $0\arcdeg >
\delta_{\rm B1950} > -30\arcdeg$, using the Parkes catalogs
\citep[PKS;][]{pkscat} and early versions of the Parkes--MIT--NRAO
catalogs \citep[PMN; since published by][]{pmn,pmnt,pmne,pmnz}.  For
the PKS sources, we required $S_{\rm 2.7~GHz} > 150$~mJy and computed
spectral indices between 0.4~GHz and 2.7~GHz, or between 2.7~GHz and
5.0~GHz, depending on the information available in the Parkes
catalogs.  For the PMN sources, we required $S_{\rm 4.9~GHz} >
120$~mJy and computed spectral indices between the 4.9~GHz PMN
measurement and the 2.7~GHz Parkes measurement.  These flat-spectrum
targets, 719~in total, typically had positions with a
two-dimensional rms accuracy of $16\arcsec$.  Confusion from Galactic
emission, particularly when spectral indices involved a 0.4~GHz
measurement, led to a zone of avoidance around the Galactic plane.

We used the VLA \citep{tho80} to observe these sources for a total of
63 hours, in four separate sessions during the period 1994~February
20--24~UT.  At this time, the array was being changed from its most
compact configuration (D) to its most extended configuration (A).  For
our observations most of the of the antennas were in their
A-configuration positions, but a few antennas (no more than 7) were
still in their D-configuration positions.

Data were acquired in dual circular polarizations at a center
frequency of 8.43~GHz and with a bandwidth of 25~MHz.  We did not use
the full 50~MHz bandwidth that is routinely available, in order to
minimize bandwidth smearing of the target sources.  (For sources as
bright as our target sources, the error in the derived positions is
dominated by systematic effects such as bandwidth smearing, rather
than statistical error.)  Observations were made assuming a coordinate
equinox of 2000.

Our phase calibrators for these observations were sources for which
positions had been measured with very-long-baseline interferometry
(VLBI) in the early 1990s, with an accuracy of approximately 2~mas
(see Russell et al.\ 1994, and references therein).  The phase
calibrators that we used, and the VLBI positions we assumed for them,
are listed in Table~\ref{tbl:phasecals}.  This table excludes the
sources that we intended to use as phase calibrators but that proved
to be too resolved for this purpose, as described in
\S~\ref{sec:data-reduction}.

We observed an additional 26 sources drawn from \cite{rus94}, but did
not use these 26 sources as phase calibrators, in order to quantify
the astrometric accuracy of our observations.  These ``astrometric
check sources'' were treated exactly the same as target sources during
the observations and data reduction.  A list of the astrometric check
sources, and the VLBI-based positions we assumed for them, is given in
Table~\ref{tbl:checksrcs}.

All sources in our program (targets, calibration sources, and
astrometric check sources) were scheduled for 2-minute observations.
The switching time between phase calibrator observations was
6~minutes, with switching angles that were usually less than
$10\arcdeg$ (but occasionally up to $15\arcdeg$).  After allowing time
for telescope slewing, and electronics settling, this provided
1--1.5~minutes of integration time per observation per source.
Additional observations of the primary flux density calibrator 3C286
(=~J1331$+$3030) were used to set the flux density scale to an
accuracy of approximately 3\%.  Information on all Stokes parameters
was obtained, but this paper presents the results for Stokes $I\/$
only.

\section{DATA REDUCTION}
\label{sec:data-reduction}

The data were calibrated with the NRAO {\sc aips}\footnote{The
Astronomical Image Processing System ({\sc aips}) is developed and
distributed by the NRAO.} software, using standard procedures.  At
this stage we recognized that some of our intended phase calibrators
had structure that was significantly resolved by the VLA.  These poor
phase calibrators do not appear in Table~\ref{tbl:phasecals}.  We also
excluded from further analysis all targets that were meant to have
been calibrated by these poor phase calibrators.

As an internal consistency check, the positions and structures of the
targets were determined using two completely different methods,
described below.  We stress that, in both methods, the centroid
positions for the targets were derived prior to any self--calibration.

The first method, identical to that employed by \citet{jvas1},
\citet{jvas2}, and \citet{jvas3} for JVAS, used an automatic procedure
in {\sc aips} that located the target within a wide--field
($82\arcsec$) map and determined its position by fitting a quadratic
profile to the brightest component in the map.  The source structure
was then determined by creating a smaller map ($20\arcsec$) and
performing two iterations of {\sc clean} and phase-only
self-calibration.  The synthesized beam, using uniform weighting, was
typically $0\farcs4 \times 0\farcs2$.  The peak flux density in the
final map was recorded, along with the total flux density within a
$10\arcsec \times 10\arcsec$ region centered on the peak.

The second method, similar to that employed by \citet{mye02} and
\citet{thesis} for gravitational lens surveys with the VLA, used an
automatic procedure in {\sc difmap} \citep{difmap} that located the
target within a wide--field map ($128\arcsec$) and determined its
position by fitting a point--source model to the visibility data in
Fourier space.  The source structure was then determined by creating a
smaller map ($25\arcsec$) and performing iterations of {\sc clean} and
phase-only self-calibration until prescribed limits on dynamic range
or map noise had been achieved.  The synthesized beam, using uniform
weighting, was typically $0\farcs4 \times 0\farcs2$.  The peak flux
density in the final map, and the total flux density in the {\sc
clean} model, were recorded.

We also inspected a plot of visibility amplitude {\em versus\/}
baseline length for each target, and discarded targets for which the
data appeared suspicious, whether because of gross resolution,
bandwidth smearing, confusion, or obviously corrupted data.  All
surviving targets were then filtered for adequate brightness ($S_{\rm
8.4~GHz}^{\rm peak} > 50$~mJy~beam$^{-1}$) and compactness ($S_{\rm
8.4~GHz}^{\rm total} < 1.2 \times S_{\rm 8.4~GHz}^{\rm peak}$).  For
those bright and compact targets, the rms differences between the
coordinates derived using the two different data-analysis methods were
8~mas in right ascension and 10~mas in declination.  There were four
extreme outliers with two--dimensional positions differing by more
than 50~mas; those four targets were discarded.

\begin{figure}
\figurenum{1}
\plotone{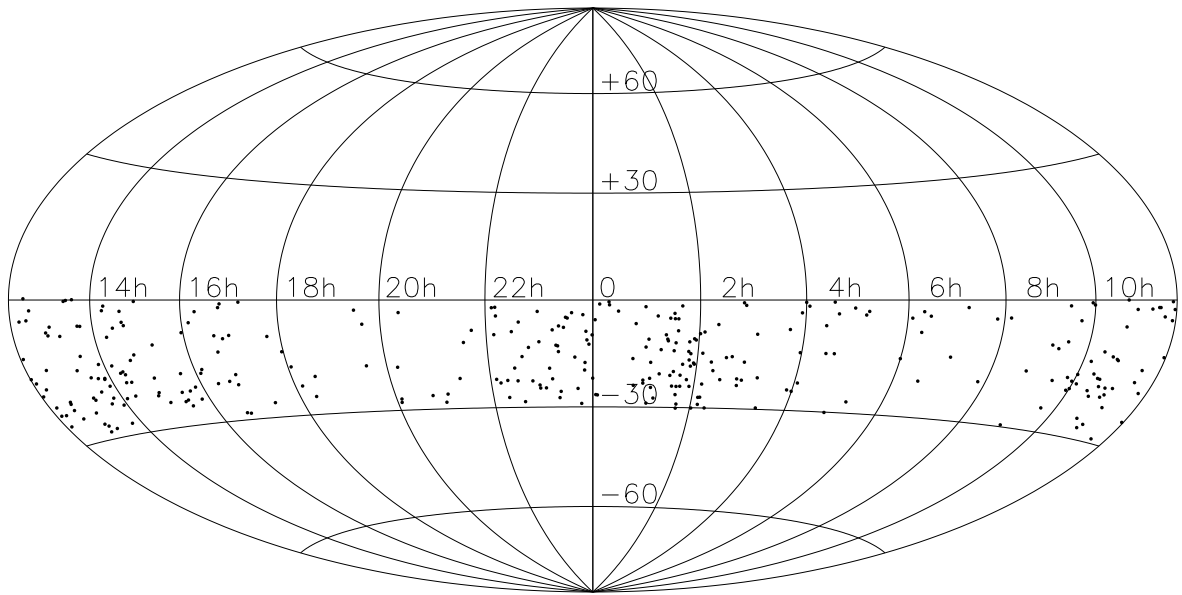}
\caption{ Celestial coordinates of the 321~compact radio sources
given in Table~\ref{tbl:catalog}, shown on an Aitoff equal-area
projection. \label{fig:aitoff} }
\end{figure}

After all of these quality--control filters had been applied, a total
of 321~bright and compact targets remained.  The locations of these
targets are shown in Fig.~\ref{fig:aitoff}, in which the Galactic zone
of avoidance is evident.  The positions and flux densities of these
321~targets are presented in Table~\ref{tbl:catalog}.  The
tabulated entries are those derived from the {\sc difmap}--based
analysis.

\begin{figure}
\figurenum{2}
\plotone{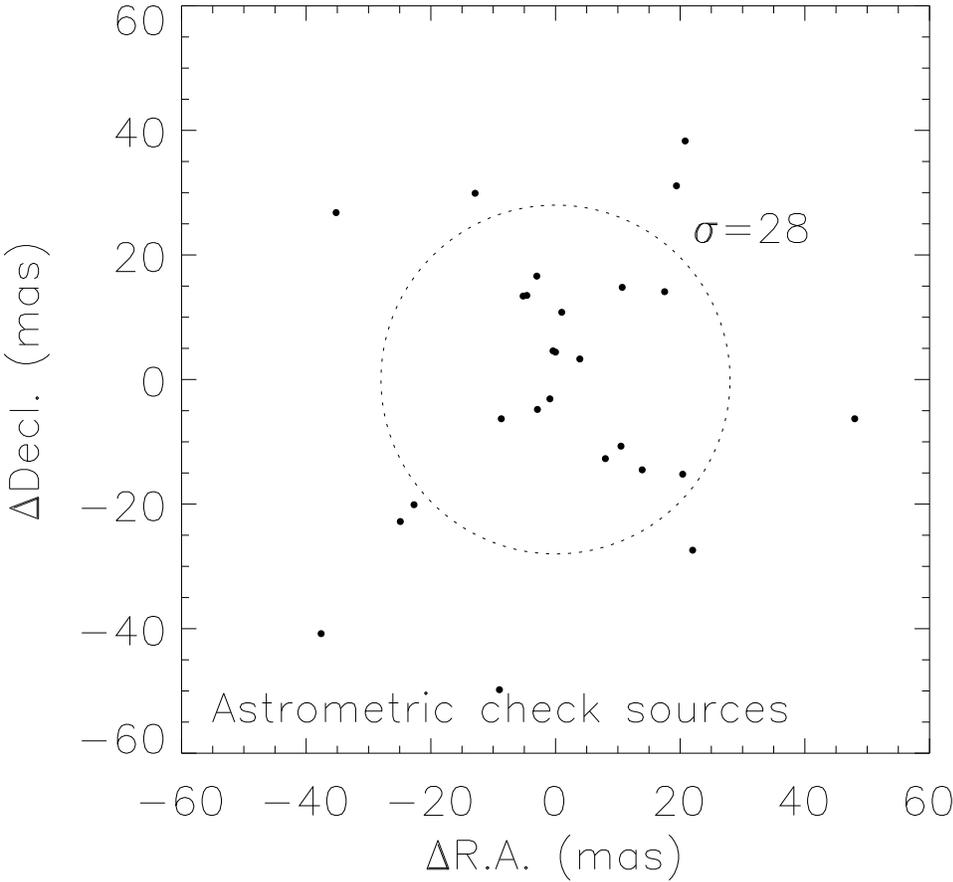}
\caption{ Differences between the positions derived from our VLA
observations and the VLBI-based positions for the 26 astrometric check
sources.  The two-dimensional rms of 28~mas is indicated by the dotted
circle. \label{fig:checksrcs} }
\end{figure}

As an external test of astrometric accuracy, we measured the positions
of the astrometric check sources using exactly the same {\sc
difmap}--based analysis as used for the target sources.  The rms
differences between our VLA-based coordinates and the VLBI-based
coordinates are 19~mas in right ascension and 21~mas in declination,
implying a two--dimensional rms accuracy of 28~mas.  The differences
between the VLBI positions and the VLA positions are plotted in
Fig.~\ref{fig:checksrcs}.

\begin{figure}
\figurenum{3}
\plotone{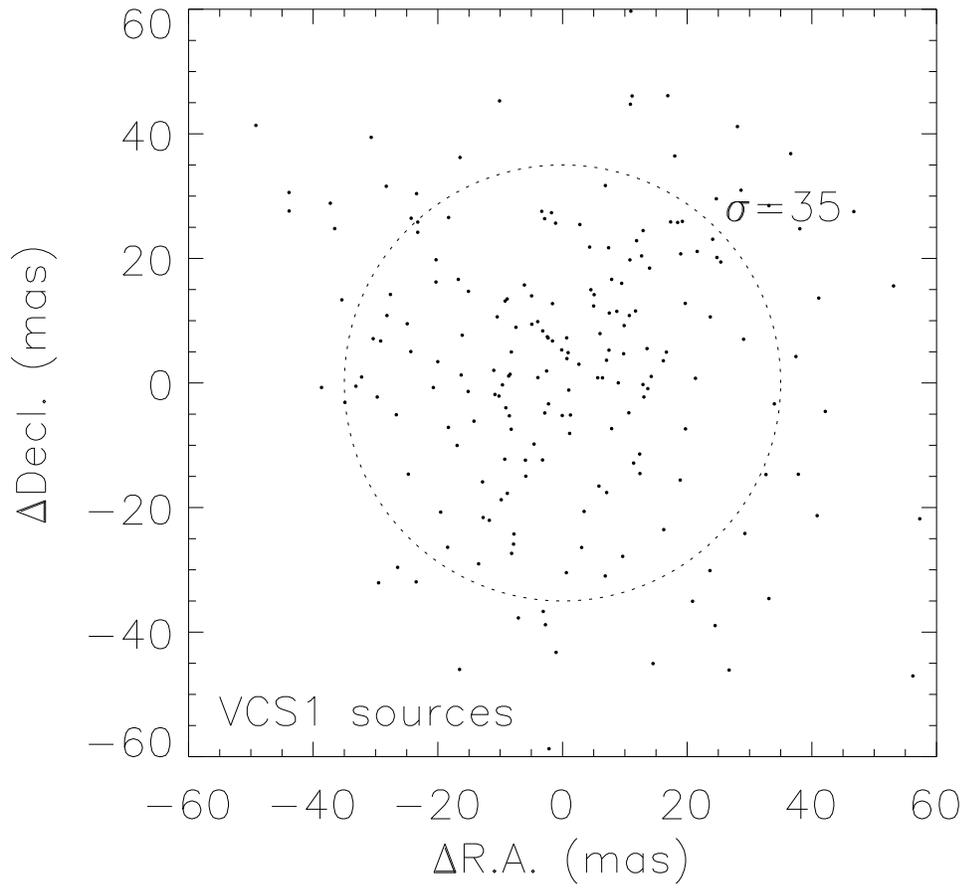}
\caption{ Differences between the positions derived from our VLA
observations and the VCS1 positions for the 201 sources in common
between our catalog and VCS1.  The two-dimensional rms of 35~mas is
indicated by the dotted circle.  Ten sources are missing from this
plot because one or both coordinate differences exceed 60~mas.
\label{fig:vcs} }
\end{figure}

A second external test of astrometric accuracy has recently become
possible, because many of the targets in a preliminary version of our
catalog were subsequently observed with the Very Long Baseline Array
(VLBA) as part of the VLBA Calibrator Survey
\citep[VCS;][]{beasley02}.  The median one-dimensional uncertainty in
the VCS1 positions (along the major axis of the error ellipse) is
0.9~mas.  The VCS1 catalog contains 201 sources in common with
Table~\ref{tbl:catalog} of this paper.  These sources are identified
by the comment ``VCS1'' in Table~\ref{tbl:catalog}.  The rms
differences between our VLA-based coordinates and the VCS1 coordinates
are 22~mas in right ascension and 27~mas in declination, implying a
two-dimensional rms accuracy of 35~mas.  The differences between the
VLBI positions and the VLA positions are plotted in
Fig.~\ref{fig:vcs}.  Based on this analysis, we estimate the
two-dimensional rms error of our catalog to be 35~mas.

\section{THE CATALOG}
\label{sec:catalog}

The positions and flux densities of the 321~new phase calibrators
found from our VLA observations are given in Table~\ref{tbl:catalog}.
These entries are derived from the {\sc difmap}--based analysis,
rather than the {\sc aips}--based analysis, although, as mentioned in
the previous section, the agreement between the two methods was
excellent.  The columns of Table~\ref{tbl:catalog} contain the
following information:
\vskip 0.1in

\noindent Col.\ 1.\ Source name derived from J2000 coordinates,\\
Col.\ 2.\ Right ascension (hours, minutes, seconds) in J2000 
          coordinates,\\
Col.\ 3.\ Declination (degrees, arcminutes, arcseconds) in J2000 
          coordinates,\\
Col.\ 4.\ Total flux density $S_{\rm 8.4~GHz}^{\rm total}$ in mJy,\\
Col.\ 5.\ Peak flux density $S_{\rm 8.4~GHz}^{\rm peak}$
          in mJy~beam$^{-1}$,\\
Col.\ 6.\ Comment.\\

The comment ``VCS1'' in the sixth column indicates that the source was
subsequently observed by \citet{beasley02} and appears in the VLBA
Calibrator Survey--1 catalog.  The positions in the VCS1 catalog
generally have a much smaller uncertainty ($\sim 1$~mas), and are
preferable to our positions when selecting a phase calibrator for VLBI
observations.

The catalog contains a few sources with declinations slightly larger
than $0\arcdeg$ because the selection was originally performed in
B1950 coordinates.  It is important to note that the catalog is not a
complete sample of compact sources in the stated declination range.
While we did observe a statistically complete sample of flat--spectrum
sources, the catalog presents only the subset of sources that survived
our very conservative quality--control filters.

In the paper version of this article, only the first page of the table
is reproduced, as an example.  The entire table is available in the
electronic version of this paper.

\section{SUMMARY AND DISCUSSION}
\label{sec:summary}

We have produced a new catalog of compact radio sources in a dense
grid in the southern sky.  The catalog is intended to be useful for
selecting phase calibration sources for radio interferometers, which
is why we took special care in estimating the astrometric accuracy of
the positions, and in discarding data that were suspect in any way.
These compact radio sources can be used as phase calibration sources
and also as positional references for proper--motion studies of nearby
objects.

Compact sources can also be utilized in other ways.  First, sources
with intrinsically small angular sizes are required to study
scintillation, angular broadening, and Faraday rotation due to the
interstellar or interplanetary medium \citep[see, e.g.,][for
reviews]{ric90,ric01}.  Second, some compact sources exhibit
intra--day variability, a topic of intense recent interest \citep[see,
e.g.,][]{jau01}.  Third, those few compact sources that have stable
flux densities, such as gigahertz--peaked--spectrum sources (GPS; see,
e.g., O'Dea~1998) and especially compact symmetric objects
\citep{fas01}, can be used as convenient flux density standards.
These sources are rare, but the compact sources in our catalog can be
investigated for this purpose.

Finally, the observing methods described in this article---snapshot
surveys of hundreds of flat--spectrum sources---have also proven
useful for finding gravitational lenses.  Lenses are fairly easy to
recognize in these surveys, because they have multiple compact
components separated by $\sim 0\farcs5$--$6\arcsec$, unlike the great
majority of flat--spectrum sources.  This technique was used by JVAS
\citep{kin99} and CLASS \citep{mye02,bro02} to find dozens of new
cases of multiple--image gravitational lensing.  Moreover,
\citet{thesis} has used the data described in this paper, along with
other VLA data, to discover 4 new lenses to date (Winn et al.\ 2000,
2001, 2002a, 2002b).  The details of that lens survey, and the
reduction of those other VLA data, will be described in a future
paper.  The subset of the observations presented here was designed
especially for astrometric accuracy, unlike those other observations,
and thus merited separate treatment.

\acknowledgments We thank I.\ Browne and P.\ Wilkinson for a
productive collaboration on JVAS in the northern sky and for useful
discussions.  We are also grateful to J.\ Russell for providing
updated positions for VLBI sources in advance of our observations; to
S.\ Myers and C.\ Fassnacht for assistance with {\sc difmap}; and to
D.\ Rusin for comments on the manuscript.  J.N.W.\ is supported by an
Astronomy \& Astrophysics Postdoctoral Fellowship, under NSF grant
AST-0104347.

\begin{deluxetable}{rrr}
\tabletypesize{\scriptsize}
\tablecaption{
Phase calibration sources and their assumed positions.
\label{tbl:phasecals}
}
\tablewidth{0pt}
\tablehead{ & Right Ascension & Declination \\ 
 J2000 Name & (J2000)         & (J2000) }
\startdata
J0006$-$0623 & 00:06:13.8929   & $-$06:23:35.334  \\
J0011$-$2612 & 00:11:01.2467   & $-$26:12:33.376  \\
J0016$-$0015 & 00:16:11.0885   & $-$00:15:12.444  \\
J0050$-$0929 & 00:50:41.3174   & $-$09:29:05.209  \\
J0113$+$0222 & 01:13:43.1449   & $+$02:22:17.317  \\
J0116$-$1136 & 01:16:12.5220   & $-$11:36:15.433  \\
J0120$-$2701 & 01:20:31.6634   & $-$27:01:24.651  \\
J0137$-$2430 & 01:37:38.3463   & $-$24:30:53.884  \\
J0141$-$0928 & 01:41:25.8321   & $-$09:28:43.673  \\
J0204$-$1701 & 02:04:57.6743   & $-$17:01:19.839  \\
J0217$+$0144 & 02:17:48.9548   & $+$01:44:49.700  \\
J0240$-$2309 & 02:40:08.1745   & $-$23:09:15.731  \\
J0241$-$0815 & 02:41:04.7984   & $-$08:15:20.752  \\
J0339$-$0146 & 03:39:30.9378   & $-$01:46:35.803  \\
J0340$-$2119 & 03:40:35.6079   & $-$21:19:31.170  \\
J0416$-$1851 & 04:16:36.5445   & $-$18:51:08.339  \\
J0423$-$0120 & 04:23:15.8007   & $-$01:20:33.064  \\
J0442$-$0017 & 04:42:38.6608   & $-$00:17:43.418  \\
J0453$-$2807 & 04:53:14.6469   & $-$28:07:37.326  \\
J0457$-$2324 & 04:57:03.1792   & $-$23:24:52.019  \\
J0501$-$0159 & 05:01:12.8099   & $-$01:59:14.255  \\
J0530$-$2503 & 05:30:07.9628   & $-$25:03:29.898  \\
J0539$-$1550 & 05:39:32.0101   & $-$15:50:30.319  \\
J0607$-$0834 & 06:07:59.6992   & $-$08:34:49.977  \\
J0730$-$1141 & 07:30:19.1124   & $-$11:41:12.599  \\
J0739$+$0137 & 07:39:18.0339   & $+$01:37:04.619  \\
J0745$-$0044 & 07:45:54.0823   & $-$00:44:17.539  \\
J0811$+$0146 & 08:11:26.7073   & $+$01:46:52.221  \\
J0826$-$2230 & 08:26:01.5730   & $-$22:30:27.203  \\
J0902$-$1415 & 09:02:16.8309   & $-$14:15:30.874  \\
J0909$+$0121 & 09:09:10.0916   & $+$01:21:35.618  \\
J0921$-$2618 & 09:21:29.3538   & $-$26:18:43.385  \\
J1024$-$0052 & 10:24:29.5865   & $-$00:52:55.499  \\
J1024$-$0052 & 10:24:29.5866   & $-$00:52:55.499  \\
J1035$-$2011 & 10:35:02.1553   & $-$20:11:34.359  \\
J1037$-$2934 & 10:37:16.0797   & $-$29:34:02.812  \\
J1048$-$1909 & 10:48:06.6204   & $-$19:09:35.726  \\
J1058$+$0133 & 10:58:29.6052   & $+$01:33:58.824  \\
J1127$-$1857 & 11:27:04.3924   & $-$18:57:17.440  \\
J1130$-$1449 & 11:30:07.0526   & $-$14:49:27.387  \\
J1133$+$0040 & 11:33:20.0558   & $+$00:40:52.838  \\
J1146$-$2447 & 11:46:08.1034   & $-$24:47:32.896  \\
J1147$-$0724 & 11:47:51.5540   & $-$07:24:41.140  \\
J1246$-$0730 & 12:46:04.2321   & $-$07:30:46.574  \\
J1246$-$2547 & 12:46:46.8020   & $-$25:47:49.287  \\
J1256$-$0547 & 12:56:11.1665   & $-$05:47:21.524  \\
J1305$-$1033 & 13:05:33.0150   & $-$10:33:19.427  \\
J1316$-$3338 & 13:16:07.9860   & $-$33:38:59.171  \\
J1337$-$1257 & 13:37:39.7828   & $-$12:57:24.692  \\
J1357$-$1527 & 13:57:11.2450   & $-$15:27:28.785  \\
J1408$-$0752 & 14:08:56.4812   & $-$07:52:26.665  \\
J1438$-$2204 & 14:38:09.4694   & $-$22:04:54.746  \\
J1448$-$1620 & 14:48:15.0542   & $-$16:20:24.548  \\
J1512$-$0905 & 15:12:50.5329   & $-$09:05:59.828  \\
J1522$-$2730 & 15:22:37.6760   & $-$27:30:10.784  \\
J1557$-$0001 & 15:57:51.4340   & $-$00:01:50.413  \\
J1625$-$2527 & 16:25:46.8916   & $-$25:27:38.326  \\
J1700$-$2610 & 17:00:53.1541   & $-$26:10:51.724  \\
J1709$-$1728 & 17:09:34.3454   & $-$17:28:53.364  \\
J1733$-$1304 & 17:33:02.7058   & $-$13:04:49.547  \\
J1743$-$0350 & 17:43:58.8561   & $-$03:50:04.616  \\
J1820$-$2528 & 18:20:57.8509   & $-$25:28:12.587  \\
J1911$-$2006 & 19:11:09.6529   & $-$20:06:55.108  \\
J1939$-$1525 & 19:39:26.6577   & $-$15:25:43.057  \\
J1939$-$1002 & 19:39:57.2562   & $-$10:02:41.519  \\
J2000$-$1748 & 20:00:57.0904   & $-$17:48:57.672  \\
J2040$-$2507 & 20:40:08.7729   & $-$25:07:46.662  \\
J2129$-$1538 & 21:29:12.1759   & $-$15:38:41.040  \\
J2134$-$0153 & 21:34:10.3096   & $-$01:53:17.238  \\
J2158$-$1501 & 21:58:06.2819   & $-$15:01:09.327  \\
J2236$-$1433 & 22:36:34.0871   & $-$14:33:22.188  \\
J2255$-$0844 & 22:55:04.2398   & $-$08:44:04.021  \\
J2258$-$2758 & 22:58:05.9629   & $-$27:58:21.255  \\
J2323$-$0317 & 23:23:31.9537   & $-$03:17:05.023  \\
J2331$-$1556 & 23:31:38.6525   & $-$15:56:57.008  \\
J2333$-$2343 & 23:33:55.2379   & $-$23:43:40.657  \\
J2337$-$0230 & 23:37:57.3391   & $-$02:30:57.628  \\
J2348$-$1631 & 23:48:02.6085   & $-$16:31:12.021  \\
J2358$-$1020 & 23:58:10.8824   & $-$10:20:08.610  \\

\enddata
\end{deluxetable}

\begin{deluxetable}{rrr}
\tabletypesize{\scriptsize}
\tablecaption{
Astrometric check sources and their assumed positions.
\label{tbl:checksrcs}
}
\tablewidth{0pt}
\tablehead{ & Right Ascension & Declination \\ 
 J2000 Name & (J2000)         & (J2000) }
\startdata
J0115$-$0127 & 01:15:17.09996  & $-$01:27:04.5762 \\
J0437$-$1844 & 04:37:01.48273  & $-$18:44:48.6118 \\
J0513$-$2159 & 05:13:49.11433  & $-$21:59:16.0898 \\
J0539$-$2839 & 05:39:54.28146  & $-$28:39:55.9462 \\
J0541$-$0541 & 05:41:38.08340  & $-$05:41:49.4281 \\
J0609$-$1542 & 06:09:40.94956  & $-$15:42:40.6709 \\
J0650$-$1637 & 06:50:24.58188  & $-$16:37:39.7239 \\
J0725$-$0054 & 07:25:50.63993  & $-$00:54:56.5436 \\
J0808$-$0751 & 08:08:15.53602  & $-$07:51:09.8851 \\
J0836$-$2016 & 08:36:39.21532  & $-$20:16:59.5033 \\
J0927$-$2034 & 09:27:51.82431  & $-$20:34:51.2308 \\
J1150$-$0023 & 11:50:43.87078  & $-$00:23:54.2037 \\
J1354$-$0206 & 13:54:06.89530  & $-$02:06:03.1896 \\
J1404$-$0130 & 14:04:45.89571  & $-$01:30:21.9475 \\
J1405$+$0415 & 14:05:01.11983  & $+$04:15:35.8204 \\
J1432$-$1801 & 14:32:57.69057  & $-$18:01:35.2468 \\
J1507$-$1652 & 15:07:04.78693  & $-$16:52:30.2655 \\
J1513$-$1012 & 15:13:44.89339  & $-$10:12:00.2633 \\
J1517$-$2422 & 15:17:41.81320  & $-$24:22:19.4754 \\
J1626$-$2951 & 16:26:06.02092  & $-$29:51:26.9702 \\
J2011$-$1546 & 20:11:15.71093  & $-$15:46:40.2523 \\
J2131$-$1207 & 21:31:35.26173  & $-$12:07:04.7957 \\
J2146$-$1525 & 21:46:22.97935  & $-$15:25:43.8843 \\
J2229$-$0832 & 22:29:40.08431  & $-$08:32:54.4342 \\
J2246$-$1206 & 22:46:18.23198  & $-$12:06:51.2769 \\
J2354$-$1513 & 23:54:30.19516  & $-$15:13:11.2114 \\

\enddata
\end{deluxetable}

\clearpage
\begin{deluxetable}{rrrrrr}
\tabletypesize{\scriptsize}
\tablecaption{
(Abridged) Catalog of compact radio sources measured at 8.4~GHz.
\label{tbl:catalog}
}
\tablewidth{0pt}
\tablehead{ & Right Ascension & Declination & 
  $S_{\rm 8.4~GHz}^{\rm total}$  & $S_{\rm 8.4~GHz}^{\rm peak}$ & \\
  J2000 Name & (J2000) & (J2000) & (mJy) & (mJy~beam$^{-1}$)    & Comment }
\startdata
J0001$-$1551 & 00:01:05.3268     & $-$15:51:07.065    & 293.4   & 289.7   & VCS1       \\
J0003$-$1927 & 00:03:18.6764     & $-$19:27:22.342    & 244.4   & 235.5   & VCS1       \\
J0004$-$1148 & 00:04:04.9139     & $-$11:48:58.378    & 645.7   & 629.5   & VCS1       \\
J0005$-$1648 & 00:05:17.9312     & $-$16:48:04.650    & 294.6   & 291.4   & VCS1       \\
J0008$-$2339 & 00:08:00.3710     & $-$23:39:18.125    & 383.3   & 379.5   & VCS1       \\
J0008$-$2559 & 00:08:26.2527     & $-$25:59:11.538    & 377.0   & 373.9   & VCS1       \\
J0010$-$2157 & 00:10:53.6520     & $-$21:57:04.179    & 362.4   & 354.2   & VCS1       \\
J0013$-$0423 & 00:13:54.1308     & $-$04:23:52.287    & 251.3   & 249.4   & VCS1       \\
J0015$-$1812 & 00:15:02.4937     & $-$18:12:50.913    & 387.3   & 379.8   & VCS1       \\
J0016$-$2343 & 00:16:05.7393     & $-$23:43:52.161    & 103.3   & 99.3    &            \\
J0022$+$0014 & 00:22:25.4257     & $+$00:14:56.164    & 625.2   & 597.7   &            \\
J0024$-$0412 & 00:24:45.9816     & $-$04:12:01.522    & 324.9   & 306.7   & VCS1       \\
J0030$-$0211 & 00:30:31.8233     & $-$02:11:56.122    & 184.6   & 180.7   &            \\
J0031$-$1426 & 00:31:56.4109     & $-$14:26:19.366    & 156.1   & 151.9   &            \\
J0034$-$2303 & 00:34:54.7960     & $-$23:03:35.221    & 126.7   & 125.1   &            \\
J0037$-$2145 & 00:37:14.8247     & $-$21:45:24.681    & 173.7   & 172.8   &            \\
J0038$-$2459 & 00:38:14.7287     & $-$24:59:02.111    & 397.6   & 390.0   & VCS1       \\
J0038$-$2120 & 00:38:29.9540     & $-$21:20:03.977    & 222.7   & 205.2   & VCS1       \\
J0051$-$0650 & 00:51:08.2114     & $-$06:50:02.218    & 800.2   & 777.0   & VCS1       \\
J0053$-$0727 & 00:53:36.5168     & $-$07:27:29.604    & 172.6   & 166.2   &            \\
J0058$-$0539 & 00:58:05.0680     & $-$05:39:52.258    & 538.3   & 522.7   & VCS1       \\
J0101$-$2831 & 01:01:52.3891     & $-$28:31:20.446    & 220.3   & 216.7   & VCS1       \\
J0102$-$2646 & 01:02:56.3546     & $-$26:46:36.530    & 164.5   & 158.3   &            \\
J0104$-$2416 & 01:04:58.2064     & $-$24:16:28.444    & 240.8   & 240.1   & VCS1       \\
J0106$-$2718 & 01:06:26.0816     & $-$27:18:11.837    & 699.3   & 695.7   & VCS1       \\
J0110$-$0741 & 01:10:50.0235     & $-$07:41:41.110    & 498.6   & 496.0   & VCS1       \\
J0115$-$2804 & 01:15:23.8834     & $-$28:04:55.221    & 669.1   & 656.8   & VCS1       \\
J0117$-$1507 & 01:17:38.9287     & $-$15:07:55.023    & 76.3    & 68.3    &            \\
J0117$-$2111 & 01:17:48.7839     & $-$21:11:06.617    & 329.0   & 325.9   & VCS1       \\
J0118$-$2141 & 01:18:57.2655     & $-$21:41:30.112    & 570.5   & 541.1   & VCS1       \\
J0122$-$0018 & 01:22:13.9078     & $-$00:18:01.050    & 196.9   & 181.7   &            \\
J0125$-$0005 & 01:25:28.8418     & $-$00:05:55.925    & 1308.9  & 1250.6  & VCS1       \\
J0131$-$1211 & 01:31:12.5474     & $-$12:11:00.688    & 87.0    & 82.1    &            \\
J0132$-$1654 & 01:32:43.4824     & $-$16:54:48.459    & 1150.8  & 1056.6  &            \\
J0134$+$0003 & 01:34:12.7061     & $+$00:03:45.125    & 315.9   & 304.8   &            \\
J0135$-$2008 & 01:35:37.5123     & $-$20:08:45.848    & 735.0   & 717.0   & VCS1       \\
J0138$-$0540 & 01:38:51.8529     & $-$05:40:08.213    & 222.8   & 217.3   & VCS1       \\
J0138$-$2254 & 01:38:57.4664     & $-$22:54:47.335    & 236.6   & 226.7   & VCS1       \\
J0142$-$1714 & 01:42:23.4059     & $-$17:14:35.446    & 122.6   & 120.4   &            \\
J0142$-$0544 & 01:42:38.8785     & $-$05:44:01.518    & 186.0   & 174.7   &            \\
J0145$-$2733 & 01:45:03.3913     & $-$27:33:34.301    & 635.2   & 627.9   & VCS1       \\
J0147$-$2144 & 01:47:07.3509     & $-$21:44:42.525    & 82.6    & 77.9    &            \\
J0151$-$1732 & 01:51:06.0827     & $-$17:32:44.730    & 347.6   & 338.8   & VCS1       \\
J0151$-$1719 & 01:51:48.0489     & $-$17:19:55.070    & 125.3   & 123.7   &            \\
J0152$-$1412 & 01:52:32.0148     & $-$14:12:39.397    & 289.4   & 262.8   & VCS1       \\
J0153$-$1906 & 01:53:01.5096     & $-$19:06:56.702    & 126.4   & 123.4   &            \\
J0154$-$2329 & 01:54:46.1008     & $-$23:29:53.908    & 123.2   & 118.6   &            \\
J0206$-$2212 & 02:06:20.0700     & $-$22:12:19.615    & 214.8   & 201.7   & VCS1       \\
J0210$-$1444 & 02:10:23.1773     & $-$14:44:58.996    & 177.8   & 174.8   &            \\
J0215$-$0222 & 02:15:42.0155     & $-$02:22:56.757    & 596.7   & 581.2   & VCS1       \\
J0217$-$0820 & 02:17:02.6612     & $-$08:20:52.315    & 332.1   & 323.9   & VCS1       \\
J0217$-$0121 & 02:17:54.9976     & $-$01:21:50.724    & 94.8    & 91.6    &            \\
J0217$-$1631 & 02:17:57.2494     & $-$16:31:10.448    & 375.8   & 371.5   & VCS1       \\
J0219$-$1842 & 02:19:21.1601     & $-$18:42:38.723    & 415.9   & 410.1   & VCS1       \\
J0220$-$2151 & 02:20:35.1475     & $-$21:51:12.055    & 235.8   & 231.8   & VCS1       \\
J0222$-$1615 & 02:22:00.7252     & $-$16:15:16.522    & 434.3   & 421.9   & VCS1       \\
J0223$-$1656 & 02:23:43.7634     & $-$16:56:37.685    & 192.9   & 191.3   &            \\
J0226$-$1843 & 02:26:47.6274     & $-$18:43:39.209    & 210.1   & 200.6   &            \\
J0227$-$0621 & 02:27:44.4625     & $-$06:21:06.741    & 176.2   & 169.3   &            \\
J0239$-$1348 & 02:39:26.0230     & $-$13:48:43.345    & 188.0   & 185.2   &            \\
J0239$-$0234 & 02:39:45.4714     & $-$02:34:40.925    & 812.8   & 802.7   &            \\
J0240$-$0504 & 02:40:56.1720     & $-$05:04:42.200    & 150.1   & 145.2   &            \\
J0243$-$0550 & 02:43:12.4692     & $-$05:50:55.295    & 607.0   & 604.9   & VCS1       \\
J0246$-$1236 & 02:46:58.4706     & $-$12:36:30.786    & 357.7   & 348.2   & VCS1       \\
J0252$-$2219 & 02:52:47.9544     & $-$22:19:25.421    & 406.1   & 398.0   & VCS1       \\
J0256$-$2137 & 02:56:12.8401     & $-$21:37:29.099    & 267.1   & 244.9   & VCS1       \\
J0259$-$0019 & 02:59:28.5152     & $-$00:19:59.981    & 824.7   & 812.3   & VCS1       \\
J0301$-$1812 & 03:01:06.7171     & $-$18:12:17.844    & 101.6   & 100.9   &            \\
J0303$-$2407 & 03:03:26.5007     & $-$24:07:11.464    & 210.1   & 186.7   &            \\
J0315$-$1031 & 03:15:56.8729     & $-$10:31:39.409    & 192.9   & 182.9   &            \\
J0325$-$2415 & 03:25:13.3422     & $-$24:15:48.053    & 302.4   & 301.2   & VCS1       \\
J0327$-$2202 & 03:27:59.9234     & $-$22:02:06.403    & 515.1   & 485.2   & VCS1       \\
J0329$-$2357 & 03:29:54.0746     & $-$23:57:08.789    & 511.1   & 501.7   & VCS1       \\
J0331$-$2524 & 03:31:08.9185     & $-$25:24:43.300    & 175.0   & 174.2   &            \\
J0339$-$1736 & 03:39:13.7048     & $-$17:36:00.785    & 69.9    & 66.8    &            \\
J0340$-$2152 & 03:40:03.4097     & $-$21:52:01.387    & 103.3   & 88.1    &            \\
J0349$-$2401 & 03:49:15.3891     & $-$24:01:14.313    & 90.5    & 85.5    &            \\
J0349$-$2102 & 03:49:57.8272     & $-$21:02:47.730    & 259.0   & 252.9   & VCS1       \\
J0357$-$0751 & 03:57:43.2936     & $-$07:51:14.555    & 329.6   & 324.8   & VCS1
\enddata
\end{deluxetable}

\end{document}